# Automated Software Testing Using Metahurestic Technique Based on An Ant Colony Optimization

Praveen Ranjan Srivastava
Computer Science and Information System Group
Birla Institute of Technology and Science,(BITS), Pilani
Rajasthan, India -333031
praveenrsrivastava@gmail.com

Km Baby
Network System Software Technology Group
Cisco Systems, Bangalore
India- 560 087
bprajapa@cisco.com

*Abstract*— **Software testing is an important and valuable part of the software development life cycle. Due to time, cost and other circumstances, exhaustive testing is not feasible that's why there is a need to automate the software testing process. Testing effectiveness can be achieved by the State Transition Testing (STT) which is commonly used in real time, embedded and web-based type of software systems. Aim of the current paper is to present an algorithm by applying an ant colony optimization technique, for generation of optimal and minimal test sequences for behavior specification of software. Present paper approach generates test sequence in order to obtain the complete software coverage. This paper also discusses the comparison between two metaheuristic techniques (Genetic Algorithm and Ant Colony optimization) for transition based testing.**

## I. INTRODUCTION

Software products should be reliable, correct, and scalable. To ensure these qualities, it is necessary to test the software at various conditions and hence software testing is an important component of the software development process [1, 2, 3]. A primary purpose of testing is to detect software errors/problems so that the defects may be discovered and corrected [1] before delivery the software. Due to the time and cost constraints, it is not possible to test the software manually and fix the defects [2]. Thus the use of test automation plays a very important role in the software testing process [3]. Now a day, Artificial Intelligence (AI) techniques are changing the nature of the test automation process [4]. It has been identified that one of the software engineering areas with a more suitable and realistic use of artificial intelligence techniques is software testing [4, 5, 6, 7] and these techniques are know as a metaheuristic approach [8]

This paper purposes an algorithm which uses an ACO optimization technique to generate the automatic state – transition test sequence, which gives a strong level of software coverage.
An ACO algorithm [9] [10] is a probabilistic technique for solving computational problems which can be used to find "good" paths through the graph. It is depends on the behavior of ants in finding paths from their colony to food [11] [12]. Using ACO to generate test sequences for state-based software testing is already presented in [13] but the main problem is complete software coverage. As the present market is highly competitive, it is a pressing need of software organizations to provide good quality software products to the customer within the estimated budget, and hence a strong level of testing coverage technique is essential.
A GA-based test data generation technique has been proposed [14] [15] to generate test data from UML state diagram, so that test data can be generated before coding, the main problem of this approach is fitness function and their chromosome length, in this case coverage of all transitions are not possible, even increase the chromosome length, hence GA based approach is not suitable of strong level software coverage [14]. GADGET [16] and TGEN [17] use genetic algorithm to improve the quality of generating test data only not any kind of coverage.
In another work [18] test data generation using genetic algorithm is presented but complete software coverage was not guaranteed.
The ACO as said earlier is a probabilistic technique for solving computational problems which can be reduced to finding good paths through graphs.
State based testing using an ACO has been nicely represented [13], this approach is well suited for state based testing ,not of all transition based coverage; still this approach is quite better for doing research using ACO in the area of software testing.

This paper is structured as follows. Section 2 presents the GA approach for test data. Section 3 represents proposed approach to test sequence generation. Section 4 describes the analysis of the suggested approach for test sequence generation, and finally in Section 5 concludes the paper.

## II. GENETIC ALGORITHM AND SOFTWARE TESTING

It has been identified that one of the Software Engineering areas with a more prolific use of AI techniques is software testing. The focus of these techniques includes the applications of genetic algorithms (GA). GA provides a general method for a searching technique, which uses the concept of natural evolution [19]. It is inspired by natural genetics to provide the solutions to problems. Genetic algorithms are inspired by Darwin's theory of evolution [20]. GA is very important for various reasons and it has been applied to a large number of scientific and engineering problems, such as automatic test data generation, optimization, machine learning, automatic programming, transportation problems, adaptive control, etc. [21][22].

GA evolves as a number of suitable test data sets, each test data for at least each path. The best test data for each parallel path are picked and used for calculating the overall coverage test data [23]. Another work by [14] [15] enhances the coverage of transitions but does not guarantee for full coverage of the test data.

In research paper [14], one of the real case studies (telephone system) is mentioned and this is illustrated in Figure 1. According to figure 1, a list of all transition is given with corresponding states and transitions. Through Genetic Algorithm [14], they find out some level of software coverage, however, the maximum coverage for the telephone system case study is not increased. In Figure 1 the transitions which are not covered in the telephone system are t7, t8, t9, t10, t11, and t12 [14].

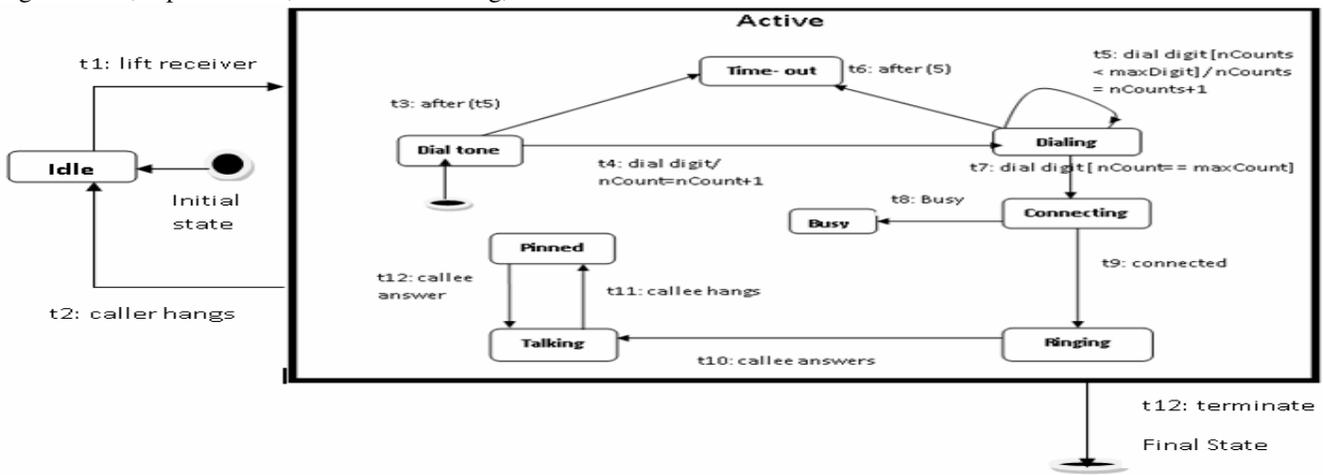

Figure 1: The telephone system state machine diagram [14]

Therefore, we can conclude[14][23] , that the GA approach is not feasible for the type of control, embedded and web based kind of software systems because the GA based proposed approach does not provide a full coverage of all the transitions.

If the software is not fully covered, there may be uncovered transitions which have defects, which in turn can cause problems in the system. Graphical representation of the above analysis [14] [23] is shown in Figure 2
:

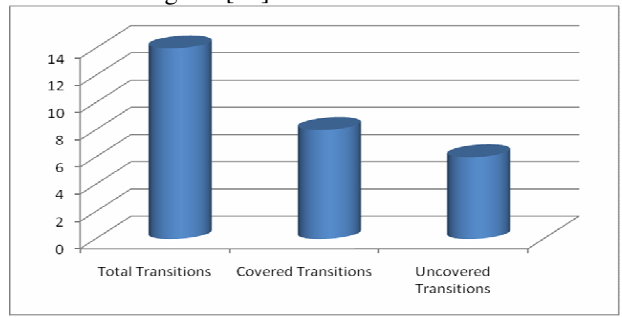

Figure 2: Transition coverage graph for telephone system

It is clear from the Figure 2 that around 43% of the transitions are uncovered. These uncovered transitions may reveal a lot of uncovered defects in the large state transition systems. Therefore, this approach using GA algorithm is not suitable for real time and control systems.

In this type of requirement, an Ant's behavior is very useful. The next section discusses about how ACO can generate full coverage of transitions in a state transition based software testing.

## III. PROPOSED APPROACH

This paper suggests the approach to generate the automatic test cases and provides a solution to cover all the transition at least once. The purpose of the ACO algorithm is to cover the optimal transition at least once in the UML [24] State Transition Diagram of the software under test. It provides the optimal test sequence of transition in the state transition diagram. Selection of transitions depends upon the probability of the transition. Higher probability values means that the chances to select the transition are also high. The probability value of transition depends upon: the feasibility of transition ($F_{ij}$), which shows direct connection between the vertices; pheromone value ($\tau_{ij}$), which helps other ants make decisions in the future, and heuristic information ($\eta_{ij}$) of the transition, which indicates the visibility of a transition for an ant at the current vertex. In some cases if there are equal probabilities of feasible transition, then by the following three steps the algorithm selects the feasible transition [12].

1. An ant will select a self-transition if it exists at a current vertex or else the ant will approach to rule 2. An ant will select the next state according to the value of visited status parameter ($V_s$). If the current vertex $V_1$ is direct connected to the vertex say $V_2$ and is not visited yet by the ant, then the ant will select $V_2$ as the next state which means that the transition ($V_1 \rightarrow V_2$) traversed. This concept fulfills the criteria of all state coverage at least once.
2. After all the above consideration, if the selection is not possible then the ant will select any feasible transition randomly.

In the proposed algorithm the ant has the ability to collect the knowledge of all feasible transitions from its current state. An approach for the feasibility check of the transitions from the current state is used. This approach is defined in the feasibility set of transition ($F_{ij}$). The ant also has four other information about transition viz.

Pheromone level on transition ($\tau_{ij}$), Heuristic information for the transitions ($\eta_{ij}$), visited states with the help of visited status ($V_s$) and the probability parameter P. After the selection of a particular transition the ant will update the pheromone level as well as the heuristic value. Pheromone level is increased according to last pheromone level and heuristic information but heuristic information, is updated only on the basis of the previous heuristic information. An ant p at a vertex 'i' (here vertex means state of state transition diagram) and another vertex 'j' which is directly connected to "i" means that there is a transition between the vertices 'i' and 'j' i.e. (i→j). In the graph this transition is associated with five tuple $F_{ij}$ (p), $\tau_{ij}$ (p), $\eta_{ij}$ (p), $V_s$(p) and $P_{ij}$ (p) where (p) shows that values of tuple associated with ant p. All description about these attribute is given below:

1. Feasible transition set: F = {$F_{ij}$ (p)} represents the direct connection with the current vertex 'i' to the neighboring vertices "j". Direct connection shows that the vertices are adjacent to the current vertex 'i', i.e. a direct edge exists in between the current vertexes 'i' and the chosen vertex 'j'.
   - $F_{ij}$ =1 means that transition between the vertex 'i' and 'j' is feasible.
   - $F_{ij}$=0 means the transition between the vertex 'i' and 'j' is not feasible.

2. Pheromone trace set: $\tau = \tau_{ij}$ (p) represents the pheromone level on the feasible transition (i→j) from current vertex 'i' to next vertex 'j'. The pheromone level is updated after the particular transition. This pheromone helps other ants to make decisions in the future.

3. Heuristic set: $\eta = \eta_{ij}$ (p) indicates the visibility of a transition, for an ant at current vertex 'i', to vertex 'j'.

4. Visited status set: Vs shows information about all the states which are already traversed by the ant p. For any state 'i':
   - $V_s$ (i) =0 shows that vertex 'i' is not visited yet by the ant p.
   - Whereas $V_s$ (i) =1 indicates that state 'i' is already visited by the ant p.

5. Probability set: Selection of transition depends on the probabilistic value of transition. Since it is inspired by the ant behavior, probability value of the transition depends on the feasibility of transition $F_{ij}$ (p), pheromone value $\tau_{ij}$ (p) and heuristic information $\eta_{ij}$ (p) of transition for ant p. There are α and β, two more parameters which are used to calculate the probability of a transition. These parameters α and β control the desirability versus visibility factors. α and β are associated with the pheromone and heuristic value of the transitions respectively. The proposed ant colony algorithm helps to not only get knowledge of the present state but also all the feasible transitions from the current state to the next state and the historical knowledge of the already traversed transitions and states by the ant.

Algorithm for ant p:
Initialize all parameter
   *1.1* Set heuristic Value (η): for every transition in the state transition diagram, heuristic value η =2.
   *1.2* Set pheromone level (τ): for every transition in state transition diagram, initialize pheromone value τ =1.
   *1.3* Set visited status ($V_s$): for every state $V_s$=0 (initially no state is visited by the ant).
   *1.4* Set Probability (P): for each transition initialize probability P=0.
   *1.5* α=1, β= 1, here α and β are the parameter which controls the desirability versus visibility i.e. desirability implies whether an ant wants to traverse any particular transition on the basis of pheromone value or not and visibility means the solution which the ant has on the basis of prior experience regarding the path. These parameters are associated with pheromone and heuristic values of the transition respectively.

**1.6** Set count: count= cyclomatic complexity.

**1.7.** Set key: key1=end _node, it is a variable which store the value of end node.

**2.** While (count>0)
 Evaluation at vertex 'i'
   **2.1**. Initialize:  i = start.
   **2.2**. Update the track:  Update the visited status for the current vertex 'i'   i.e. if ($V_s[i] ==0$) then $V_s[i] =1$
   **2.3** Evaluate Feasible Set: Means to determine F (p) for the current vertex 'i'. This procedure evaluates the entire possible transition from the current vertex 'i' to the all the neighboring vertices with the help of a state transition diagram. If there is no feasible path then go to step 3.0.

   **2.4** Sense the trace: To sense the trace, evaluate the probability from the current vertex 'i' to all the non-zero connections in the F (p). As discussed earlier the ant's behavior is probabilistic. For every non-zero element belonging to the feasible set F (p), we calculate the probability with the help of the below formula.

$$P_{ij} = \frac{(\tau_{ij})^{\alpha} * (\eta_{ij})^{-\beta}}{\sum_{1}^{k} ((\tau_{ik})^{\alpha} * (\eta_{ik})^{-\beta})}$$

for every k belonging to feasible set F (p).

 **2.5**. Move to next vertex: Using the rule below to move to the next vertex
  R1: Select paths (i→j) with maximum probability ($P_{ij}$).
  R2: If two or more paths (from i→j and i→k) have equal probability like ($P_{ij}$  =  $P_{ik}$) then select path according to below rule:
       R2.1. If there is self transition i.e. (i→i) then select it.
       R2.2. Compare each entry in the feasible set with the end_ node i.e.
If (j== end _node) then select
 'k' as the next node otherwise follow R2.3
       R2.3. If $V_s[j] =V_s[k]$ then select randomly

 **2.6**. start= next_node

 **2.7**. Update the parameter:
   2.7.1 Update Pheromone: Pheromone is updated for path (i→j) according to the following rule
  If (j== end_node && $\tau_{ij} == \infty$)
      goto step 2.
  else if (j== end_node)
      $\tau_{ij} == \infty$
  else ($\tau_{ij}$) = ($\tau_{ij}$)$^{\alpha}$ + ($\eta_{ij}$)$^{-\beta}$
   2.7.2 Update Heuristic:
  $\eta_{ij} = 2*(\eta_{ij})$

 **2.8**. if (start! = end_node)
           goto step 2.3.
   else   count =count-1.
           goto step 2

 **2.9** End //end of algorithm
Count represents the cyclomatic complexity of a method and as count becomes zero; it shows all the decision nodes traversed. The algorithm will stop automatically in two conditions, firstly if there is no feasible transitions from the current node and secondly if the all feasible transitions are covered at least once.

## IV.   ANALYSIS OF PROPOSED ALGORITHM

In proposed algorithm comparison can be done in three areas: - (1) uses of this approach in the real world, (2) advantages of the previous work and (3) advantages of the existing approach in the field of software testing.
Software industry now days focuses more on web-based, real time, embedded, scientific, objects oriented and agent based software systems. In any real time software systems there are transitions from one state to another as in system specification. We can say that the software field is based on the object behavior and its complete transitions; this shows the need of testing approach to test various transitions, objects states or pages etc for any real kind of software. In present approach, the tester can easily get the automated optimal test sequence generation in the state transition diagram of software under test. Therefore the resultant is the increase in the probability of finding the error, within specific constraints.
ACO approach has been already used in such kind of problem [13], but the main limitation of the work for any real software[13], states are not only important ,including states, transitions are also equal importance ,but in [13] only states are captured., not much worried about transitions. This paper tries to compare the previous work [25] with the proposed modified algorithm. We apply STTACO tool [25], GA [14] and the proposed approach in one of the real time case studies which is shown in Figure 3. The figure explains the state chart diagram of an Enrollment system which was used by [14] [23].

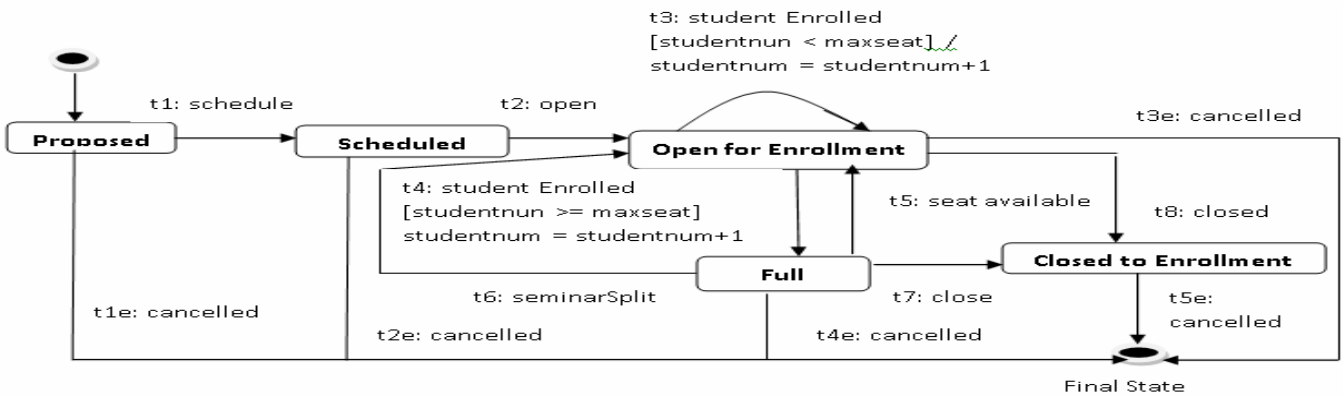

Figure 3: the enrollment state machine diagram [14]

Figure 3 explains the enrolment system; in this regard an earlier approach used GA [14] to cover for transition based coverage.

The limitation of approach [14] is the facts that were not looked for full coverage, and also looping problem is yet another concern which indicates the drawback of previous approached. In above enrolment system using GA we can achieve a maximum coverage at 64.29 percent [14] whereas the suggested approach guaranteed for full coverage i.e. 100% coverage (as shown in figure 4). The execution of the proposed approach is shown below which provides the full coverage automated test sequence. As ants have the dynamic knowledge of the system it helps them to make full coverage.

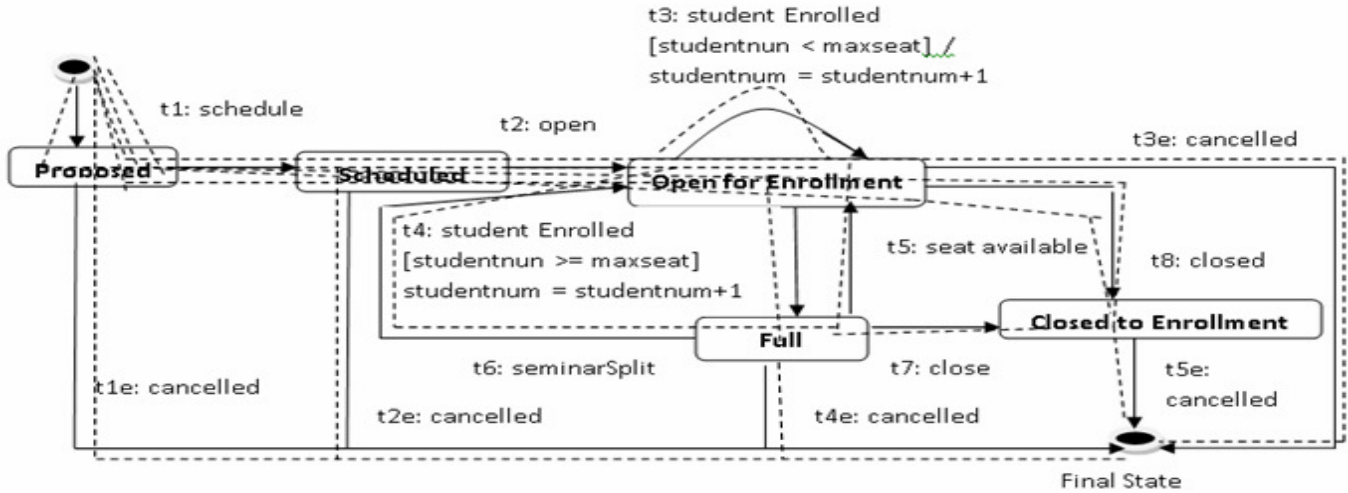

Figure 4: Optimal solutions by Proposed ACO approach for enrollment state machine

Test sequence for transitions covered which are represented with dotted lines as shown in figure 5 are: Proposed=s0, schedule=s1, open enrolment=s2, full=s3, close enrolment=s4 and sequence of test data generated by proposed approach is as follows:
Start→s0→s1→s2→s2→s3→s4→final→start→s0→final→start→s0→s1→final→start→s0→s1→s2→s4→start→s0→s1→s2→final→start→s0→s1→s2→s3→s2→s4→start→s0→s1→s2→s3→final.

In the test sequence s0→final, s1→final, s3→final, s4→final, s2→final will not repeat as these transitions are blocked by $\tau_{ij} = = \infty$. Due to this value of the pheromone, the probability came out as zero which ensured no further repetition of the transitions. Few states and transitions are also repeated in the above test sequence and the reason is that the class enrolment system is the example of a real world application where someone cannot go for open enrollment without passing through the proposed schedule and one can go for cancelation after the proposed, schedule, open for enrollment, if course is full and if enrolment is closed. In each case there is particular order and therefore some states and transitions repeat in the test sequence. The proposed approach minimizes a countable number of transitions. Taking a simple case, the above software covered by STTACO[25] tool shows that transitions s0→final, s1→final, s3→final, s4→final, s2→final may repeat many times which may lead to repeat of other sequence of transitions because the algorithm stops each time when it counts infeasible states or the final node. But in the proposed approach s0→final, s1→final, s3→final, s4→final, s2→final transitions are blocked after traversing once which further avoids the repetition of other ambiguous sequence of transitions.

Figure 5 shows that the proposed approach improved (in terms of repetitions only) the previous work [25] with full coverage. This figure also represents the relative strength of transition coverage by proposed tool and the work done by GA [14] [23].Figure 6 represents the strength of the two metaheuristic techniques ACO (proposed method) and GA [14] which indicate the guarantee for fully coverage of software under test using the proposed method.

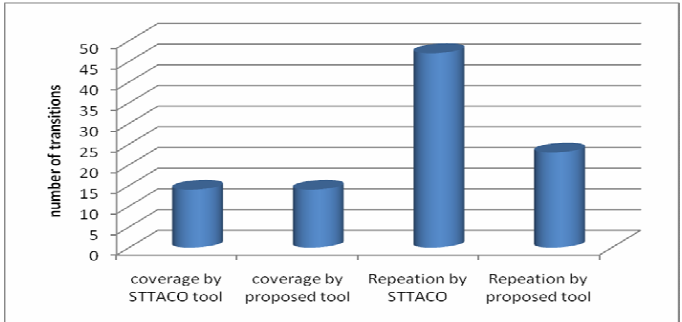

Figure 5: Relative coverage and repetition of proposed tool with STTACO tool

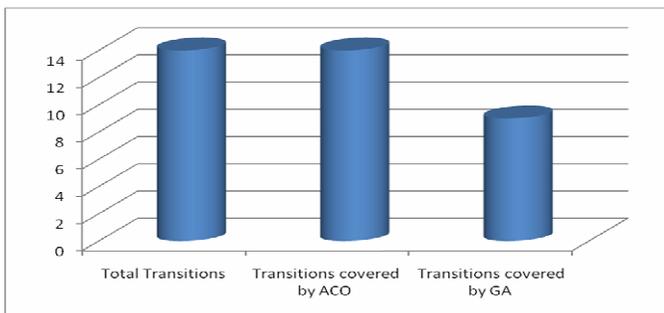

Figure 6: Relative strength of ACO with GA

## V. CONCLUSION

This paper describes the standard method of state transition based testing and its coverage level. Since the existing method using GA does not provide full coverage for software, in this context this paper proposed a model for transition based coverage by using ant colony optimization. The results that we got by applying proposed method are very encouraging. By using the strength of the ACO approach, this paper demonstrates the generation of the optimal test sequence for the state -transition based software testing. This approach enhances the tool used in [25] as it limits the repeated number of transitions in the test sequence and also provides the full coverage.

Soft computing techniques can evaluate transition based testing efficiently which may ultimately help the software industry to a greater extent. A number of extensions and applications of the model may be possible by using techniques like artificial neural networks, evolutionary computation and a combination of neuro-fuzzy approaches. These techniques can be used to model more complex nonlinear problems and in fact there is considerable need for applied research and strategy evaluation in this area using these techniques. This proposed model in this chapter is to be further extended for future work in emphasizing the difference between the ACO and other heuristic approaches (TS, GA, SA etc) used to select a path. Also future work depends on the stochastic property of transition by which we can judge which transition should be covered first.